\documentclass[pra,twocolumn,superscriptaddress]{revtex4-1}
\usepackage{graphicx,amsmath,amssymb}
\usepackage{color}
\usepackage[colorlinks=true, citecolor=blue, urlcolor=blue,citecolor=blue]{hyperref}
\begin{document}
\title{Floquet engineering to exotic topological phases in systems of cold atoms}
\author{Hui Liu}
\affiliation{School of Physical Science and Technology, Lanzhou University, Lanzhou 730000, China}
\author{Tian-Shi Xiong}
\affiliation{Department of Physics, National University of Singapore, Singapore 117542}
\author{Wei Zhang}
\affiliation{Department of Physics, Renmin University of China, Beijing 100872, China}
\affiliation{Beijing Key Laboratory of Opto-Electronic Functional Materials and Micro-Nano Devices, Renmin University of China, Beijing 100872, China}
\author{Jun-Hong An}
\email{anjhong@lzu.edu.cn}
\affiliation{School of Physical Science and Technology, Lanzhou University, Lanzhou 730000, China}
\begin{abstract}
Topological phases with a widely tunable number of edge modes have been extensively studied as a typical class of exotic states of matter with potentially important applications. Although several models have been shown to support such phases, they are not easy to realize in solid-state systems due to the complexity of various intervening factors. Inspired by the realization of synthetic spin-orbit coupling in a cold-atom system [Z. Wu {\it et al.}, Science \textbf{354}, 83 (2016)], we propose a periodic quenching scheme to realize large-topological-number phases with multiple edge modes in optical lattices. Via introducing the periodic quenching to the Raman lattice, it is found that a large number of edge modes can be induced in a controllable manner from the static topologically trivial system. Our result provides an experimentally accessible method to artificially synthesize and manipulate exotic topological phases with large topological numbers and multiple edge modes.
\end{abstract}
\maketitle

\section{Introduction} Since the discovery of the quantum Hall effect \cite{PhysRevLett.49.405}, exotic phases with topologically protected edge modes have attracted extensive attention in the past decades. The study in this field has enriched our understanding of topological nature of matters, and has led to the discovery of topological insulators \cite{RevModPhys.82.3045,PhysRevLett.98.106803}, topological superconductors \cite{RevModPhys.83.1057}, Weyl semimetals \cite{PhysRevLett.107.127205,PhysRevB.83.205101,PhysRevX.5.031013,PhysRevB.84.235126,Lu622,Xu613}, and photonic topological insulators \cite{Lu2014,nmat3520,Slobozhanyuk2016,Maczewsky2017,RevModPhys.91.015006}. An intriguing direction in this field is to seek for novel phases with large topological numbers. Such phases can provide more edge modes, which are expected to improve performance of certain devices by lowering the contact resistance in quantum anomalous Hall insulators \cite{PhysRevLett.111.136801,PhysRevLett.112.046801}. They may also be used to realize reflectionless waveguides, combiners, and one-way photonic circuits in photonic devices \cite{PhysRevLett.100.013904,PhysRevLett.113.113904,PhysRevLett.115.253901}. Although theoretical studies suggest several models that support such phases \cite{PhysRevLett.115.253901,Goldman_2009}, an experimental realization is not easy due to the complication of interplay among various types of degrees of freedom in solid-state systems.

Another possible route toward realizing topological phases is to periodically drive a traditional insulator to become a so-called Floquet topological insulators. The band structure of a given system can be drastically altered by periodic driving, such as an electromagnetic field \cite{PhysRevLett.99.047401,PhysRevB.79.081406,np1926,PhysRevB.84.235108,PhysRevLett.108.056602,PhysRevLett.105.017401} and periodic quenching \cite{PhysRevB.88.104511,PhysRevE.90.032138,PhysRevB.95.144304,PhysRevB.93.075405,PhysRevLett.115.236403,PhysRevLett.113.076403,PhysRevB.95.104308,PhysRevB.97.245430,PhysRevLett.116.176401,PhysRevLett.122.253601,2019arXiv190307678P}. In addition, periodic driving can also induce an equivalent long-range hopping which is crucial for certain exotic topological phases \cite{PhysRevB.93.144307,PhysRevB.93.184306}. However, the implementation of these schemes is usually difficult in veritable materials. Recently, the realization of synthetic spin-orbit coupling via Raman transition has paved the route toward quantum emulation of topological systems in ultracold atomic gases \cite{Lin2011,Jingzhang2016,Wu83,PhysRevLett.121.113204}. Owing to the high controllability therein, cold atoms confined in traps and optical lattices provide a promising platform to synthesize and study topological matters, as demonstrated in various examples~\cite{RevModPhys.89.011004,PhysRevX.7.031057,Trane1701207,Gross995,PhysRevLett.119.123601,PhysRevA.95.043634,PhysRevA.95.023607,ZhuSL2018,RevModPhys.91.015005,PhysRevA.91.053617}.

In this work, we investigate exotic phases with large topological numbers of the cold-atom systems confined in an optical lattice with synthetic spin-orbit coupling implemented by a Raman lattice. By periodically driving the offset phase between the Raman lattice and the optical lattice, we find that a widely tunable number of edge modes can be generated at ease in both the one- (1D) and two-dimensional (2D) cases. A criterion to determine the change of topological numbers when changing the driving parameters across the phase boundaries is established. As the experimental techniques of Raman lattice and periodic driving both have been successfully demonstrated in cold-atom systems, our proposal can be readily implemented.
\begin{figure}
 \includegraphics[width=0.9\columnwidth]{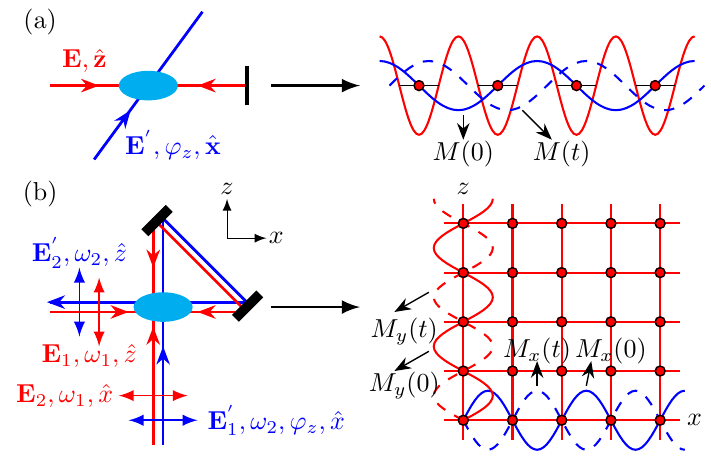}
 \caption{Scheme of periodically quenched optical lattices. An effective spin-orbit coupling of atoms with $\Lambda$-type level structure in (a) 1D and (b) 2D optical lattices is synthesized by the Raman transition driven by a standing (red) and a running wave (blue).}
\label{1}
\end{figure}

\section{Floquet topological phases}
We are interested in topological phase transitions in a time-periodic system $\hat{H}(t)=\hat{H}(t+T)$ with period $T$. The Floquet theorem promises a complete basis $|u_\alpha(t)\rangle$ from $[\hat{H}(t)-i\hbar\partial_t]|u_\alpha(t)\rangle=\varepsilon_\alpha|u_\alpha(t)\rangle$. Here $|u_\alpha(t)\rangle$ and $\varepsilon_\alpha $ play the same role as the stationary states and eigen energies in static systems. They are called quasistationary states and quasienergies \cite{PhysRevA.7.2203,PhysRevA.91.052122}. It is in the quasienergy spectrum that the topological properties of our periodic system is defined. The Floquet equation is equivalent to $\hat{U}_T|u_\alpha(0)\rangle=e^{-i\varepsilon_\alpha T/\hbar}|u_\alpha(0)\rangle$, where $\hat{U}_T=\mathbb{T}e^{-i/\hbar\int_{0}^{T}\hat{H}(t)dt}$ is the evolution operator with $\mathbb{T}$ being the time-ordering operator. Thus, $\hat{U}_T$ defines an effective static system $\hat{H}_\text{eff}\equiv {i\hbar\over T}\ln \hat{U}_T$ that shares the same (quasi)energies with the periodic system. Then one can use the well-developed tool of topological phase transition in static systems to study periodic systems via $\hat{H}_\text{eff}$.

To facilitate understanding of the underlying physics, we consider a Hamiltonian $\mathcal{H}(\mathbf{k})=\mathbf{h}(\mathbf{k})\cdot\pmb{\sigma}$ with the parameter in $\mathbf{h}$ periodically quenched between two chosen $\mathbf{h}_1$ and $\mathbf{h}_2$ within the respective time duration $T_1$ and $T_2$ \cite{PhysRevLett.106.220402}. Applying the Floquet theorem, we obtain $\mathcal{H}_\text{eff}(\mathbf{k})=\mathbf{h}_\text{eff}(\mathbf{k})\cdot \pmb{\sigma}$ \cite{PhysRevB.93.184306} with the Bloch vector $\mathbf{h}_\text{eff}(\mathbf{k})=-\arccos (\varepsilon)\underline{\mathbf{r}}/T$ and
\begin{eqnarray}
\varepsilon&=&\cos|T_1 \mathbf{h}_1(\mathbf{k})|\cos|T_2 \mathbf{h}_2(\mathbf{k})|-\underline{\mathbf{h}}_1\cdot\underline{\mathbf{h}}_2\nonumber\\ &&\times\sin|T_1\mathbf{h}_1(\mathbf{k})|\sin|T_2\mathbf{h}_2(\mathbf{k})|,\\
\mathbf{r}&=&\underline{\mathbf{h}}_1\times\underline{\mathbf{h}}_2\sin|T_1\mathbf{h}_1(\mathbf{k})|\sin|T_2\mathbf{h}_2(\mathbf{k})|-\underline{\mathbf{h}}_2\cos|T_1\mathbf{h}_1(\mathbf{k})|\nonumber\\
&&\times\sin|T_2\mathbf{h}_2(\mathbf{k})|-\underline{\mathbf{h}}_1\cos|T_2\mathbf{h}_2(\mathbf{k})|\sin|T_1\mathbf{h}_1(\mathbf{k})|,\label{varepsilon}
\end{eqnarray}
where $T=T_1+T_2$ and $\underline{\mathbf{v}}\equiv\mathbf{v}/|\mathbf{v}|$ is the unit vector of $\mathbf{v}$.
The topological properties of the system are crucially dependent on the presence or absence of the chiral symmetry, which is tunable by the periodic quenching \cite{PhysRevB.90.125143,PhysRevLett.121.036402,PhysRevLett.120.260501}. For example, the chiral symmetry is present if $\mathcal{U}_c\mathcal{H}(\mathbf{k}) \mathcal{U}_c^{-1}=-\mathcal{H}(\mathbf{k})$,
which is obviously satisfied by $\mathcal{U}_c=\sigma_\alpha$ when the $\alpha$-component of $\mathbf{h}_\text{eff}$ is absent.
Thus the chiral symmetry is present if the Bloch vector has only two components. Supplying a good way to recover the chiral symmetry by eliminating a component of $\mathbf{h}_{\text{eff}}(\mathbf{k})$, the periodic quenching can be used to generate topological phases in different classes, e.g., multiple Majorana edge modes in Kitaev chains \cite{PhysRevB.87.201109}. Note that if both of $\mathcal{H}_{j}(\mathbf{k})$ have the identical chiral symmetry and $\mathcal{U}_c$, a unitary transformation can convert $\mathcal{H}_\text{eff}(\mathbf{k})$ into the one with the same chiral symmetry (see Appendix \ref{appe1D}). 

Different from the static case, the edge modes for the periodically driven system can occur at both of the quasienergies $0$ and $\pi/T$. It means that two topological invariants which separately count the edge modes at $0$ and $\pi/T$ are generally needed \cite{PhysRevB.90.125143}. The topological phase transition is associated with the closing and reopening of the quasienergy bands. We obtain from Eq. \eqref{varepsilon} that the bands close when
\begin{eqnarray}
&&\underline{\mathbf{h}}_1=\pm \underline{\mathbf{h}}_2,\label{h1h2}\\
&&T_1|\mathbf{h}_1(\mathbf{k})|\pm T_2|\mathbf{h}_2(\mathbf{k})|=n\pi,~n\in\mathbb{Z},\label{bdt}
\end{eqnarray}
with quasienergy being zero ($\pm\pi/T$) for even (odd) $n$, or
\begin{equation}
T_j|\mathbf{h}_j(\mathbf{k})|=n_j\pi,~n_j\in\mathbb{Z}.\label{scond}
\end{equation} As the sufficient condition for judging the phase transition, Eqs. (\ref{h1h2})-(\ref{scond}) offer a guideline to design the quenching scheme to generate various topological phases at will.

\section{Cold-atom system}
Inspired by the experimental realization of synthetic spin-orbit coupling in cold-atom systems \cite{Lin2011,Jingzhang2016,Wu83,PhysRevLett.121.113204}, we consider setups in $d$ dimensions ($d=1,2$) as depicted in Fig. \ref{1}. The Hamiltonian is
\begin{equation}
\hat{H}={\mathbf{p}^2/2m}+V_{lat}(\mathbf{r})+\mathbf{M}(\mathbf{r})\cdot\pmb{\sigma}+m_z\sigma_z,\label{cdh}
\end{equation}
where $m_z$ is the Zeeman splitting, $V_{lat}(\mathbf{r})=\sum_{j=1}^dV_j \cos^2(k_0 r_j)$ is the optical lattice, and $\mathbf{M}(\mathbf{r})$ is the Raman lattice. $V_{lat}(\mathbf{r})$ is formed by a standing wave $\mathbf{E}=\hat{\bf z}E\cos (k_0x)$ for $d=1$, and by two standing waves $\mathbf{E}_{1}=\hat{\bf z}E_{1} \cos(k_0x)$ and $\mathbf{E}_{2}=\hat{\bf x}E_{2}\cos(k_0z)$ for $d=2$ of frequency $\omega_1$. $\mathbf{M}(\mathbf{r})$ is formed by the combined actions of the aforementioned standing waves and additional running waves, which take the form $\mathbf{E}'=\hat{\bf x}E'e^{i(k_0z+\varphi_{z})}$ for $d=1$, and $\mathbf{E}'_{1}=\hat{\bf x}E'_{1}e^{i(k_0z+\varphi_{z})}$ and $\mathbf{E}'_{2}=\hat{\bf z}E'_{2}e^{i(-k_0x+\varphi_{z}-\delta)}$ for $d=2$ of frequency $\omega_2$. Here, $\varphi_z$ is the initial phase and $\delta=L|\omega_2-\omega_1|/c$ with $L$ being the optical path difference. The lattice potentials $V_{lat}(\mathbf{r})$ and $\mathbf{M}(\mathbf{r})$ together induce a two-photon Raman transition between the near-degenerate ground states $|g_{\downarrow,\uparrow}\rangle$ mediated by the excited state $|e\rangle$ \cite{PhysRevLett.112.086401}. By adiabatically eliminating $|e\rangle$, we have $\mathbf{M}(\mathbf{r})=(Me^{i\varphi_{z}}\cos(k_0x),0,0)$ with $M\propto E E'$ for $d=1$, and $\mathbf{M}(\mathbf{r})=e^{i\varphi_{z}}[M_{x}\cos(k_0x)\sin(k_0z),M_{z}\cos(k_0z)\sin(k_0x),0]$ for $d=2$ with $M_{x}\propto E_{1}E'_{1}$ and $M_{z}\propto E_{2}E'_{2}$ when $\delta=\pi/2$.

Expanding Eq. \eqref{cdh} in the basis of $s$-band Wannier functions $\phi_{s\sigma}^{\bf{j}}({\bf r})$, we obtain
\begin{eqnarray}
\hat{H}&=&-\sum_{\langle \bf{i},\bf{j}\rangle,\sigma}v_0^{\bf{i}\bf{j}}\hat{c}_{\bf{i}\sigma}^{\dagger}\hat{c}_{\bf{j}\sigma}+\sum_{\langle \bf{i},\bf{j}\rangle}[v_{so}^{\bf{i}\bf{j}}\hat{c}_{\bf{i}\uparrow}^{\dagger}\hat{c}_{\bf{j}\downarrow}+\text{H.c.}]\nonumber\\
&&+\sum_{\bf i}m_z(\hat{c}_{\bf{i}\uparrow}^{\dagger}\hat{c}_{\bf{i}\uparrow}-\hat{c}_{\bf{i}\downarrow}^{\dagger}\hat{c}_{\bf{i}\downarrow}),\label{Haml}
\end{eqnarray}
where $\langle \bf{i},\bf{j}\rangle$ denotes that the summation is subjected to nearest neighbors, $v_0^{\bf{i}\bf{j}}=\int d^{d}\mathbf{r}\phi_{s\sigma}^{\bf{i}}(\mathbf{r})[{\mathbf{p}^2/2m}+V_{lat}(\mathbf{r})]\phi_{s\sigma}^{\bf{j}}(\mathbf{r})$, and $v_{so}^{\bf{i}\bf{j}}=\int d^{d}\mathbf{r}\phi_{s\uparrow}^{\bf{i}}(\mathbf{r})[\mathbf{M}(\mathbf{r})\cdot\pmb{\sigma}]\phi_{s\downarrow}^{\bf{j}}(\mathbf{r})$ \cite{Wu83}. Owing to the periodicity of the potential, we have $v_0^{\bf{i}\bf{i}\pm\bf{1}}\equiv v_0$ for both the 1D and 2D cases. Also, one can verify that $v_{so}^{\bf{i}\bf{i}\pm\bf{1}}=\pm (-1)^{i_x}v_{so}$ for $d=1$, and $v_{so}^{i_x,i_x\pm 1}=\pm (-1)^{i_x+i_z}v_{so}$ and $v_{so}^{i_z,i_z\pm 1}=\pm i (-1)^{i_x+i_z}v_{so}$ for $d=2$ \cite{PhysRevLett.110.076401,PhysRevLett.115.045303}.
Defining $\hat{c}_{\bf{j}\downarrow}\rightarrow e^{i\pi \bf{j}}\hat{c}_{\bf{j}\downarrow}$ and making the Fourier transform $\hat{c}_{\bf{i}\sigma}=\frac{1}{\sqrt{N}}\sum_{\bf{k}} \hat{c}_{\bf{k}\sigma}e^{-i\mathbf{k}\cdot\mathbf{r}}$ with $N$ being the total site number, we obtain $\hat{H}=\sum_{{\bf k}\in \text{BZ}}C^\dagger_{\bf k} \mathbf{h}(\mathbf{k})\cdot\pmb{\sigma}C_{\bf k} $ with $C^\dagger_{\bf k}=(\hat{c}^\dagger_{{\bf k}\uparrow},\hat{c}^\dagger_{{\bf k}\downarrow})$ and the summation over the first Brillouin zone (BZ). The Bloch vectors $\mathbf{h}(\mathbf{k})$ read
\begin{eqnarray}
\mathbf{h}_{1D}({ k})&=&(0,2v_{so}\sin k,m_z-2v_0\cos k),\label{blc1}\\
\mathbf{h}_{2D}({\bf k})&=&[2v_{so}\sin k_z,2v_{so}\sin k_x,\nonumber\\
&&m_z-2v_0(\cos k_x+\cos k_z)],
\end{eqnarray}
where the lattice constant has been set to one. The particle-hole symmetry is naturally kept. Because $\mathbf{h}_{1D}({ k})$ has two components, the static Hamiltonian possesses the chiral symmetry with $\mathcal{U}_c=\sigma_x$ and belongs to the symmetry class BDI \cite{RevModPhys.88.035005,PhysRevB.82.235114}. The topological properties are characterized by the winding number $\mathcal{W}=\frac{1}{2\pi}\int_{-\pi}^{\pi}dk\langle u_k|i\partial_k |u_k\rangle$ with $|u_k\rangle$ the Bloch states of the 1D Hamiltonian. When $|m_z|<2|v_0|$, the system has $\mathcal{W}=\pm 1$ and hosts one pair of edge modes \cite{PhysRevB.89.085111}. The topological property for $d=2$ is described by the Chern number $\mathcal{C}=\frac{1}{2}\sum_{\mathbf{k}\in \mathbb{D}_z} \text{sgn}[\mathbf{h}_{2D}(\mathbf{k})]_z\text{Ch}(\mathbf{k})$ for the lower band, where $\text{Ch}(\mathbf{k})=\text{sgn}[\partial_{k_x}\mathbf{h}(\mathbf{k})\times \partial_{k_z}\mathbf{h}(\mathbf{k})]_{z}$ is the chirality and $\mathbb{D}_{z}$ is the set of band-touching points for $\mathbf{h}(\mathbf{k})$ excluding the $z$ component \cite{PhysRevB.85.165456}. The system has $\mathcal{C}=\text{sgn}(m_z)$ and one pair of edge states when $|m_z|<4|v_0|$ \cite{PhysRevLett.112.086401}.

To realize the phases with larger topological numbers and more edge modes than the static cases, we consider an experimentally accessible periodic-quenching protocol
\begin{equation}
v_{so}(t) =\begin{cases}A_1 ,& t\in\lbrack mT, mT+T_1)\\A_2,& t\in\lbrack mT+T_1, (m+1)T). \end{cases}~m\in \mathbb{Z}
\label{procotol}
\end{equation}
It is achievable by either tuning the Raman lattice depth $M$ or changing the phase $\varphi_{z}$ of the running waves.

\begin{figure}[tbp]
 \includegraphics[width=.98\linewidth]{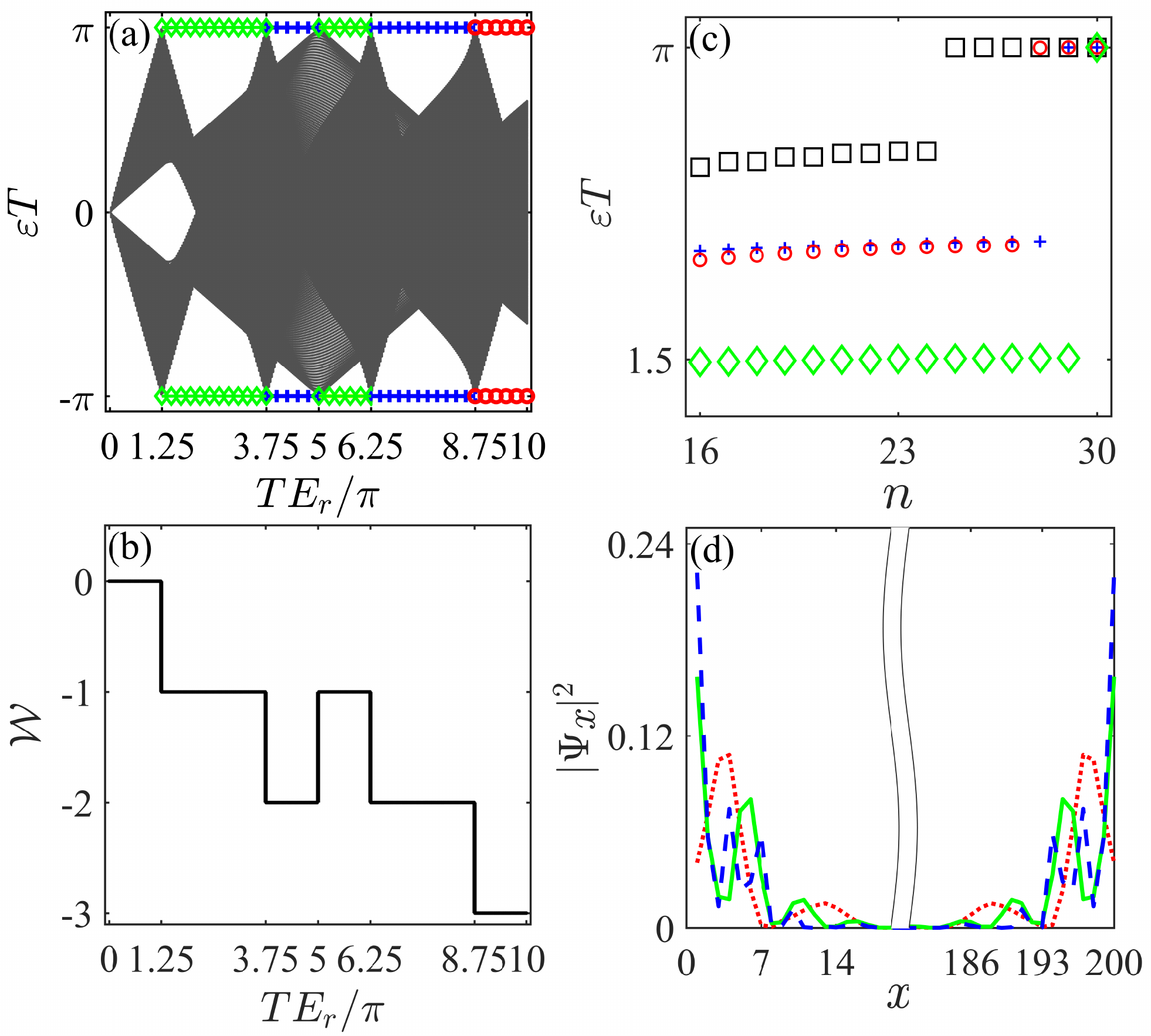}
 \caption{(a) Quasienergy spectrum and (b) winding number of the 1D system by changing $T$. (c) Quasienergies in open boundary condition when $T=2.4\pi/E_r$ (green diamond), $6.7\pi/E_r$ (blue cross), $9.2\pi/E_r$ (red circle), and $20.0\pi/E_r$ (black block). (d) Site distribution of the edge states when $T=9.2\pi/E_r$. We use $m_z=0.5E_r$, $v_0=0.15E_r$, and $v_{so}=0.4E_r$. }
 \label{2}
\end{figure}

\section{Numerical results}
\subsection{Periodic quenching in 1D case}
Choosing $T_1=T_2\equiv T/2$ and $A_2=-A_1\equiv v_{so}$ by pairwisely changing $\varphi_z$ between $0$ and $\pi$, we find that the $y$ component of $\mathbf{h}_\text{eff}(k)$ is zero. Thus $\mathcal{H}_\text{eff}(k)$ keeps the chiral symmetry and still belongs to the symmetry class BDI. From Eqs.~(\ref{h1h2}), (\ref{bdt}), and (\ref{blc1}), we have the conditions for the bands closing as follows.

\textbf{Case I}: $T|\mathbf{h}_j(\mathbf{k})|=2n\pi$. If this is satisfied for some $k$ at a certain $T$, then for any subsequent $T$ one always has a $k$ that holds the condition. Thus the bands keep closed at the zero quasienergy and no phase transition occur.

\textbf{Case II}: $\underline{\mathbf{h}}_1(\mathbf{k})=-\underline{\mathbf{h}}_2(\mathbf{k})$. Equation \eqref{h1h2} requires $k=\arccos(m_z/2v_0)$. Further with Eq. \eqref{bdt}, we obtain $n=0$. Thus, the bands always touch at the zero quasienergy irrespective of $T$ and no phase transition can take place.

\textbf{Case III}: $\underline{\mathbf{h}}_1(\mathbf{k})=\underline{\mathbf{h}}_2(\mathbf{k})$. Equation \eqref{h1h2} reveals the bands touch at $k=0$ or $\pi$. Using Eq. \eqref{bdt}, we have
\begin{equation}
n_\alpha=T|m_z- 2e^{i\alpha}v_0|/\pi
\label{k0}
\end{equation}
for $n_\alpha\in\mathbb{Z}$ and $\alpha=0,\pi$, and $\mathbf{h}_\text{eff}(\alpha)=(0,0,m_z-2e^{i\alpha}v_0)$.
The phase transition at zero quasienergy for even $n_\alpha$ requires $|m_z|<2|v_0|$ like the static case, which causes the existence of a definite $k$ such that the bands keep closed according to \textbf{Case II}. It in turn rules out the phase transition at the zero quasienergy.
Thus a phase transition occurs only for odd values of $n_\alpha$ satisfying Eq. \eqref{k0}.

Figures \ref{2}(a) and \ref{2}(b) show the quasienergy spectrum and the winding number with changing $T$. The parameters $|m_z|>2|v_0|$ are chosen such that the static systems are topologically trivial. The nontrivial phases are induced when the periodic quenching is on [see Fig. \ref{2}(a) with $E_r=\hbar^2k_0^2/2m$ being the recoil energy]. For small $T$, Eq. \eqref{scond} is not fulfilled and a finite band gap exists at the zero quasienergy. With increasing $T$, the gap remains closed due to \textbf{Case I}. However, the gap at $\pm\pi/T$ is closed and reopened at $TE_r/\pi=1.25$, $3.75$, $5.0$, $6.25$, and $8.75$ accompanied by the corresponding change of $\mathcal{W}$ [see Fig. \ref{2}(b)]. They correspond to Eq. \eqref{k0} with $n_\alpha=1_\pi$, $3_\pi$, $-1_0$, $5_\pi$, and $7_\pi$, respectively. Figure \ref{2}(c) shows that the number of degeneracy of the formed bound modes exactly equals to $|\mathcal{W}|$, as required by the bulk-edge correspondence \cite{PhysRevB.90.125143}. More bound modes are achievable with further increasing $T$. As expected, all the bound modes are highly confined at the edges [see Fig. \ref{2}(d)].

\begin{figure}[tbp]
 \includegraphics[width=0.98\linewidth]{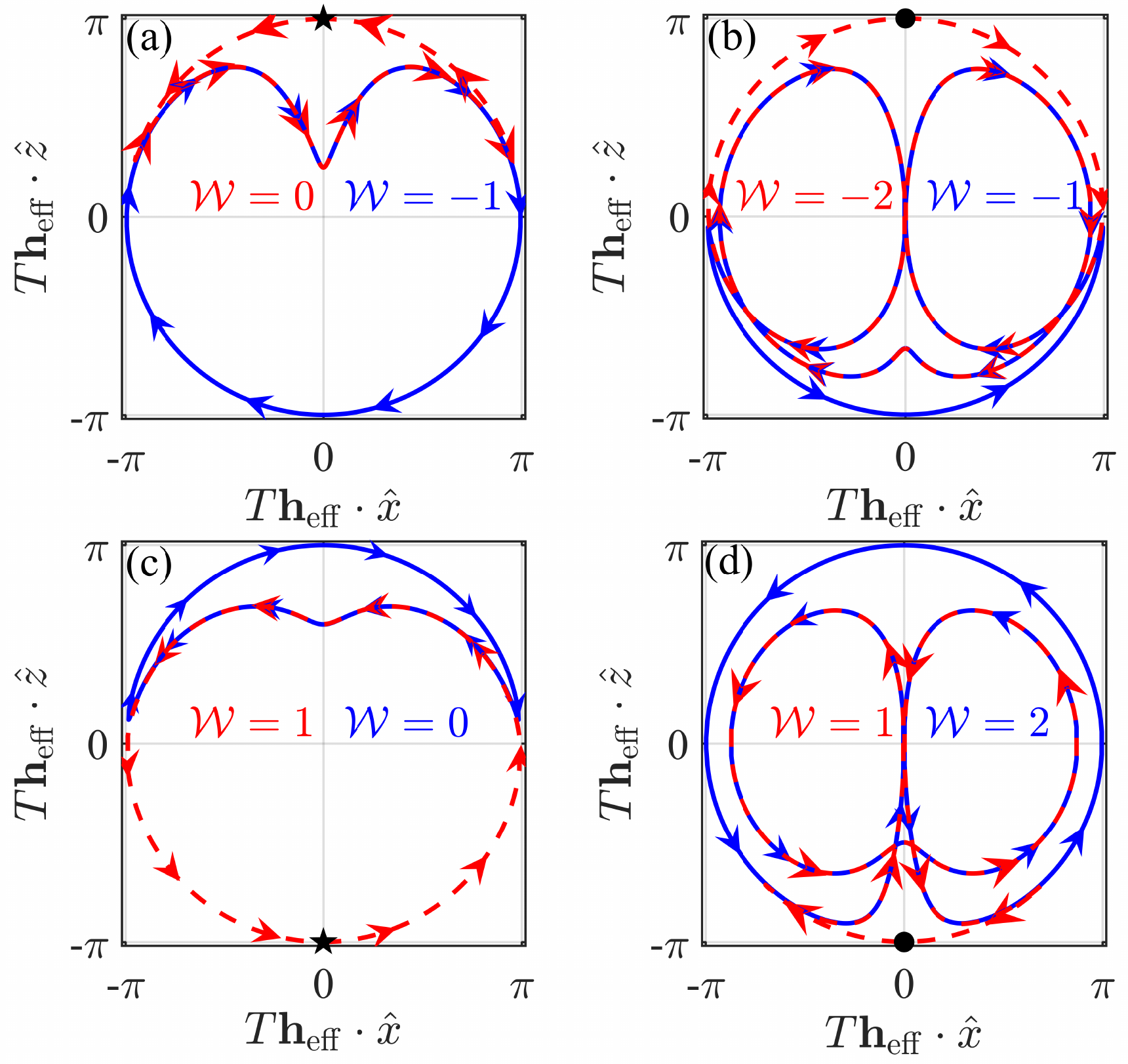}
 \caption{Trajectories of $\mathbf{h}_\text{eff}(k)$ in the 1D case for $T$ (red dashed lines) and $T+\Delta T$ (blue solid lines) crossing the boundaries of $\alpha=0$ marked by $\bullet$ and $\pi$ by $\bigstar$. The parameters satisfy $m_z-2e^{i\alpha}v_0 >0$ in panel (a) and panel (b) and $<0 $ in panel (c) and panel (d). We use $\Delta T=0.002\pi/E_r$, $v_{so}=0.05E_r$, and ($m_z/E_r, v_0/E_r,TE_r/\pi)=(0.5,0.15,1.249)$ in panel (a), $(0.6,0.16,3.571)$ in panel (b), $(-0.6,0.05,1.999)$ in panel (c), and $(-0.7,0.25,2.499)$ in panel (d). } \label{blc}
\end{figure}
To reveal how $\mathcal{W}$ changes when $T$ crosses the phase boundaries, we check the change rate of $\mathbf{h}_\text{eff}(k)$ across the quasienergy $\pm\pi/T$ at $\alpha$, i.e., $\lim_{\epsilon\rightarrow 0}\partial_k\mathbf{h}_\text{eff}(k)\big|_{k=\alpha}$ at $T_\epsilon=T-\epsilon$. The Bloch vectors near $\alpha$ read $\mathbf{h}_j(\alpha+e^{i\alpha}\delta)=[0,(-1)^j2v_{so}\delta,m_z-2e^{i\alpha}v_0]$ with $\delta$ being an infinitesimal. Then we have $T_{\epsilon}\textbf{h}_\text{eff}(\alpha+e^{i\alpha}\delta)=[{4v_{so}\delta\over m_z-2e^{i\alpha}v_0},0,\text{sgn}(m_z-2e^{i\alpha}v_0)\epsilon'_\alpha]$ with $\epsilon'_\alpha=\epsilon|m_z-2e^{i\alpha}v_0|$. Reminding that $n_\alpha$ is odd, we can easily find that the bands touch at $\text{sgn}(m_z-2e^{i\alpha}v_0)\pi/T$. The change rates of $\mathbf{h}_\text{eff}(k)$ are
\begin{eqnarray}
\lim_{\epsilon\rightarrow 0}T_\epsilon\partial_k \mathbf{h}_\text{eff}(k) |_{k=\alpha}&=&({4 e^{i\alpha}v_{so}\over m_z-2e^{i\alpha}v_0},0,0),\label{zerocon}
\end{eqnarray}
with which the change rule of $\mathcal{W}$ can be obtained.

For $ m_z-2e^{i\alpha}v_0>0$, the bands for both $\alpha=\pi$ and $0$ touch at $\pi/T$. Equation \eqref{zerocon} reveals that $\mathbf{h}_\text{eff}(k)$ crosses $\pi/T$ along the $-x$ (or $+x$) direction for $\alpha=\pi$ (or $0$) with increasing $k$. This is confirmed by the dashed lines in Figs.~\ref{blc}(a) and \ref{blc}(b). Since only the first Brillouin zone $[-\pi/T,\pi/T)$ of the quasienergy is meaningful, $\mathbf{h}_\text{eff}(k)$ abruptly jumps from $\pi/T-\delta$ to $-\pi/T+\delta$ keeping the direction unchanged when $T$ crosses the phase boundary [see the solid lines in Figs.~\ref{blc}(a) and \ref{blc}(b)]. Then a closed path with a clockwise wrapping to the origin is formed in Fig.~\ref{blc}(a) with $k$ running over $[-\pi,\pi)$. It causes $\mathcal{W}$ changing from $0$ to $-1$.
Before the phase transition, $\mathbf{h}_\text{eff}(k)$ wraps the origin twice in the clockwise direction [see Fig.~\ref{blc}(b)] indicating $\mathcal{W}=-2$. After the phase transition, an anticlockwise path is formed and $\mathcal{W}=-1$. Thus $\mathcal{W}$ decreases (or increases) $1$ with increasing $T$ across the phase boundary of $\alpha=\pi$ (or $0$). This can be confirmed by $m_z-2e^{i\alpha}v_0<0$, where the bands for both $\alpha=\pi$ and $0$ touch at $-\pi/T$. Figures~\ref{blc}(c) and \ref{blc}(d) demonstrate the cases that $\mathcal{W}$ changes from $1$ (dashed line) to $0$ (solid line) and from $1$ (dashed line) to $2$ (solid line), respectively.
\begin{figure}[tbp]
 \includegraphics[width=0.9\linewidth]{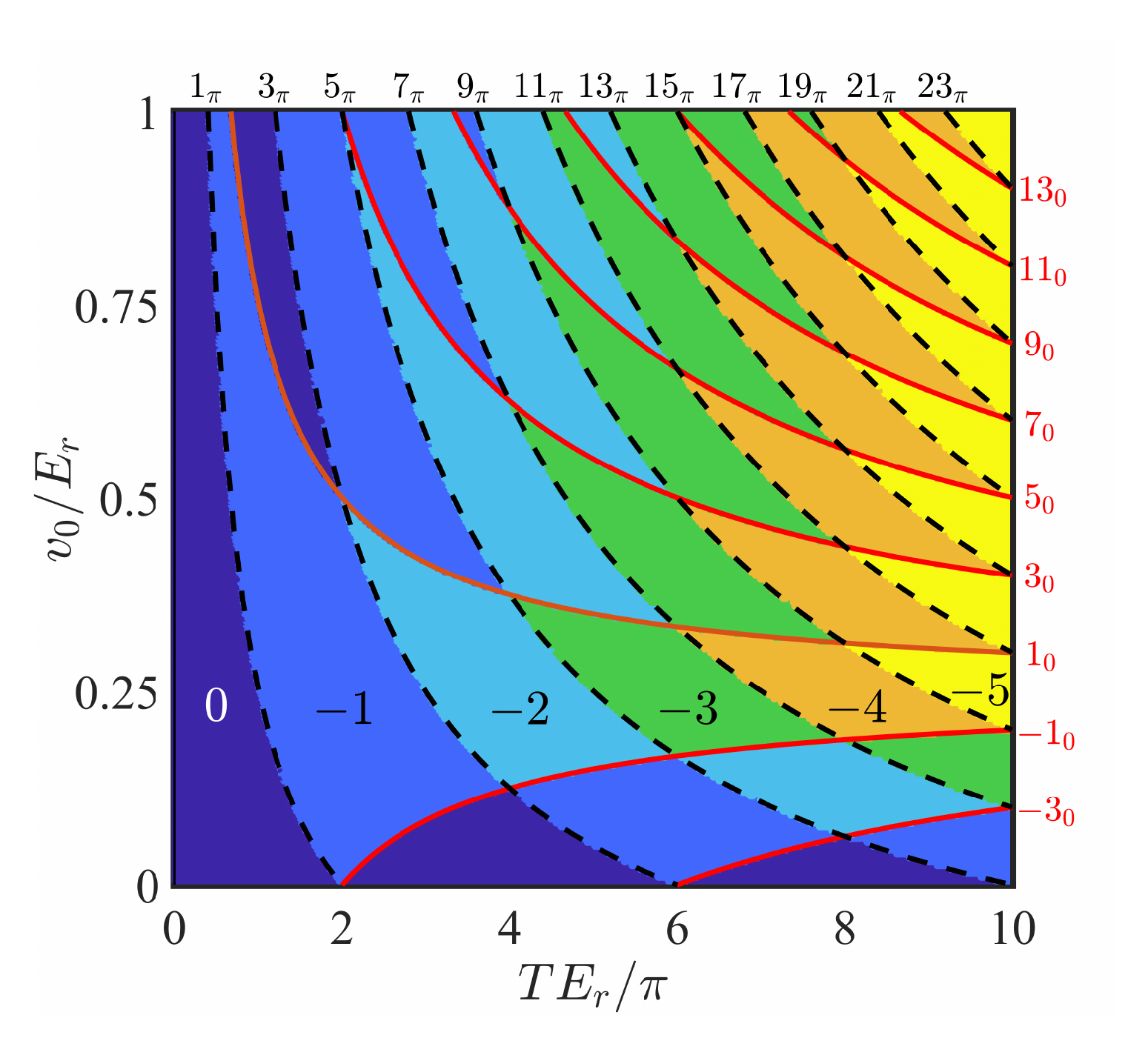}
 \caption{Phase diagram in the 1D case. The red solid and black dashed lines are phase boundaries from Eq.~\eqref{k0} with $n_\alpha$ labeled explicitly. We use $m_z=0.5E_r$ and $v_{so}=0.05E_r$.}
 \label{winding}
\end{figure}

The change of $\mathcal{W}$ can be verified by the phase diagram in Fig. \ref{winding}. The solid and dashed lines depict the phase boundaries analytically evaluated from Eq. \eqref{k0}. With increasing $T$, $\mathcal{W}$ increases $1$ through a solid line and decreases $1$ through a dashed line. Note that the phases with large $\mathcal{W}$ and multiple edge modes can be obtained at large $T$. The physics behind this originates from the ability of periodic driving in effectively engineering the long-range hopping \cite{PhysRevB.87.201109}. However, we face a tradeoff between the increased number of edge modes and the decreased
gap in the quasienergy spectrum. Considering the necessary protection of the edge modes by a nonzero bulk gap, one may not want to push our driving protocol too far. The phase diagram gives a map for experimentally designing the parameters to engineer exotic topological phases. We emphasize that the findings above are qualitatively valid in the general case with $T_1\neq T_2$ (see Appendix \ref{appe1D}).

\begin{figure}[tbp]
 \includegraphics[width=\linewidth]{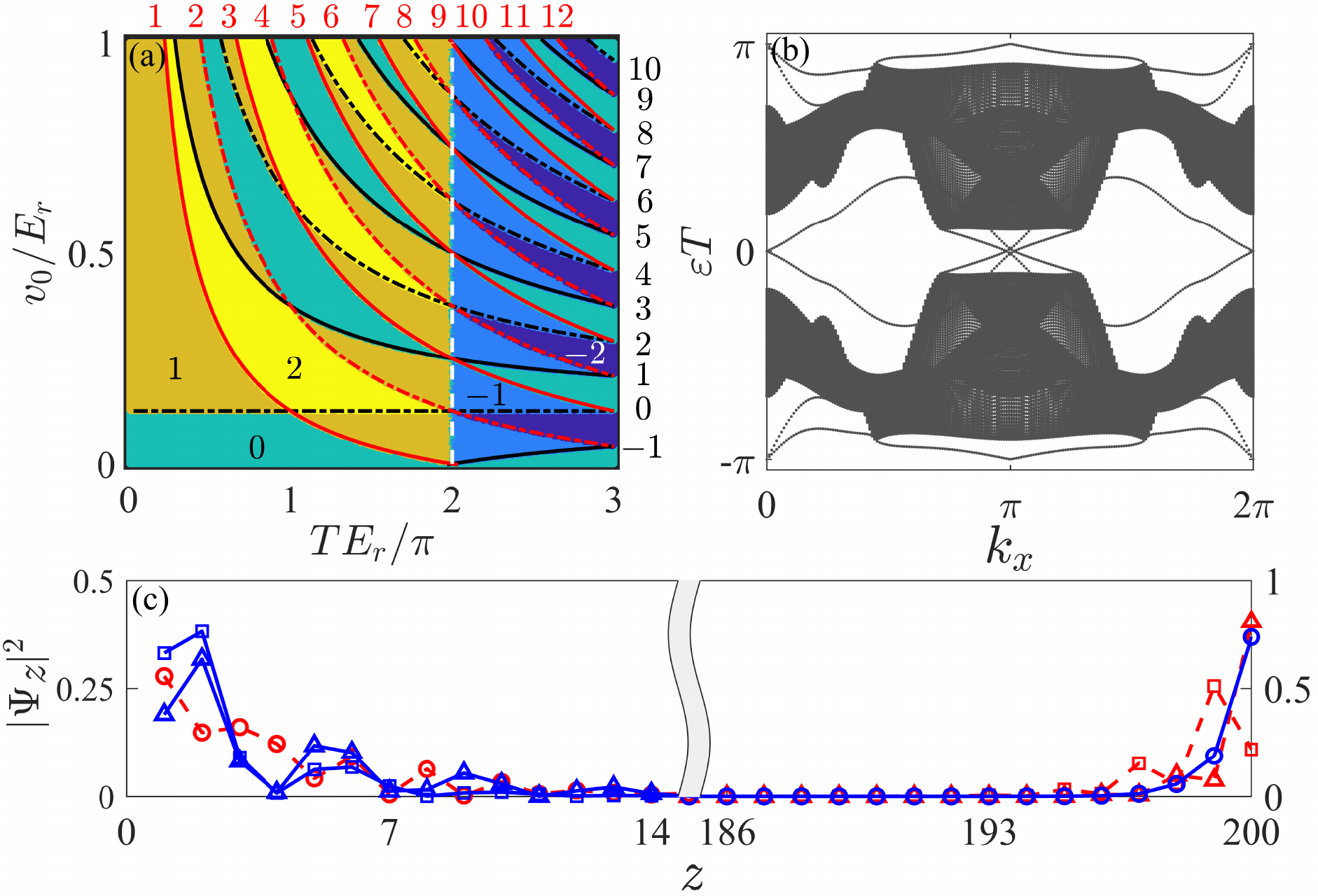}
 \caption{(a) Phase diagram in the 2D case. The black (red) lines are from Eq. (\ref{2db}) with $\alpha=\beta=0$ ($\pi$) and $n$ labeled explicitly. The white dashed line depicts the case of $(\alpha,\beta)=(0,\pi)$ or $(\pi,0)$. (b) Quasienergy spectrum and (c) site distribution of the edge states with $z$ direction opened when $(v_0/E_r,TE_r/\pi)=(0.3,2.6)$. The blue solid (red dashed) lines marked by $\triangle$, $\square$, and $\circ$ are the zero ($\pi/T$) modes. We use $m_z=0.5E_r$, $v_{so}=0.28E_r$, and $(T_1,T_2)=(0.7,0.3)T$.}
 \label{chernnm}
\end{figure}

\subsection{Periodic quenching in 2D case}
Using the protocol Eq.~\eqref{procotol}, a widely tunable number of edge modes can also be generated in the 2D case. Without loss of generality, we choose $T_1\neq T_2$. It can be proved that both of Eq. \eqref{h1h2} with ``$-$" and Eq. \eqref{scond} do not support phase transition (see Appendix \ref{appe2D}). Equation \eqref{h1h2} with ``$+$'' reveals the band-touching points $\mathbf{k}=(\alpha,\beta)$ with $\alpha,\beta=0$, or $\pi$, such that Eq. (\ref{bdt}) reads
\begin{equation}
T|m_z-2(e^{i\alpha}+e^{i\beta})v_0|=n\pi.\label{2db}
\end{equation}
Thus, the bands touch at $\pi/T$ ($0$) for odd (even) $n$.

As depicted in the phase diagram Fig.~\ref{chernnm}(a), a tunable $\mathcal{C}$ ranging from $-2$ to $2$ can be formed. The boundaries match well with Eq. \eqref{2db}. In the same mechanism, when $T$ increases across the phase boundaries, $\mathbf{h}_\text{eff}(\mathbf{k})$ has an abrupt jump in sign at $\mathbf{k}=(\alpha,\beta)$. It causes the change of the wrapping time of $\mathbf{h}_\text{eff}(\mathbf{k})$ to the origin of the BZ. The chirality $\text{Ch}(\mathbf{k})$ for $\mathbf{k}=(0,0)$ and $(\pi,\pi)$ is $-1$ (see Appendix \ref{appe2D}).
Thus when $T$ increases across the phase boundary contributed by these two points, $\Delta\mathcal{C}$ is $\text{sgn}[m_z-2v_0(e^{i\alpha}+e^{i\beta})]$ for odd $n$ and $-\text{sgn}[m_z-2v_0(e^{i\alpha}+e^{i\beta})]$ for even $n$, respectively.
 This is verified by the black and red lines in Fig. \ref{chernnm}(a). The chirality $\text{Ch}(\mathbf{k})$ for $\mathbf{k}=(0,\pi)$ and $(\pi,0)$ is $+1$. Both of the points contribute to $\Delta \mathcal{C}$ being $2\text{sgn}[(-1)^{n}m_z]$, as verified by the white dashed line in Fig. \ref{chernnm}(a).

The quasienergy spectrum in Fig. \ref{chernnm}(b) reveals that although $\mathcal{C}=-2$, there are six independent edge states. The bulk-edge correspondence can be recovered by considering the respective contribution of $\pi/T$- and zero-mode edge states to $\mathcal{C}$ \cite{PhysRevX.3.031005,PhysRevB.89.121401}. Figure \ref{chernnm}(c) shows the site distribution of the six edge states with a positive group velocity $\partial_{k_x}{\varepsilon}$, where the number of the zero and $\pi/T$ modes are both three. Two of the zero modes locate at the left edge and one at the right, which gives $\mathcal{C}_0=-1$. Two of the $\pi/T$ modes reside at the right and one at the left, which gives $\mathcal{C}_\pi=-1$ (see Appendix \ref{appe2D}). The total Chern number is $\mathcal{C}=\mathcal{C}_\pi+\mathcal{C}_0=-2$.

\section{Discussion and Conclusion}
Our result is realizable in the state-of-the-art of cold-atom experiments \cite{Lin2011,Jingzhang2016,Wu83,PhysRevLett.121.113204}. The topological phase transition in the static case of a $^{87}$Rb degenerate gas has been observed in Ref. \cite{Wu83}. The Zeeman splitting $m_z=0.1\sim0.6E_r$, the spin-conserved hopping $v_0=4.1\sim 5.0E_r$, and the spin-orbit coupling $v_{so}$ as high as $1.3E_r$ have been realized. The parameters used in our calculation are under the scope of this experimental achievement. Via pairwisely changing the phase $\varphi_z$ of the running waves between $0$ and $\pi$, no extra burden to the experiment is introduced by our periodic quenching protocol. An estimation for the $^{87}$Rb in the optical lattice formed by the red laser gives $E_r\simeq 25$kHz, which conveys from our phase diagrams in Figs. \ref{winding} and \ref{chernnm}(a) the scale of the period $T\simeq 0.2$ms. The topological numbers for our periodic system are hopefully detected by linking numbers \cite{Tarnowski2019} via observing the cyclic evolution governed by $\hat{H}(t)$ of the ground state of $\hat{H}(0)$, where several periods of driving may be generally needed.

In summary, we have proposed a periodic quenching scheme to generate exotic phases with large topological numbers and multiple edge modes both in 1D and 2D cold-atom systems. Resorting to the periodic switching of the phase of the Raman lattice between $0$ and $\pi$, our scheme can be readily implemented in experiments with existing techniques of synthesizing spin-orbit coupling~\cite{Lin2011,Jingzhang2016,Wu83,PhysRevLett.121.113204}, and supplies an avenue to controllably design topological devices in an experimentally friendly way.

\section*{Acknowledgments}
This work is supported by the National Natural Science Foundation (Grant Nos. 11875150, 11834005, 11434011, 11522436, and 11774425),
the National Key R\&D Program of China (Grant No. 2018YFA0306501), the Beijing Natural Science Foundation (Grant No. Z180013),
and the Fundamental Research Funds for the Central Universities of China.

\appendix

\begin{figure*}
 \includegraphics[width=2\columnwidth]{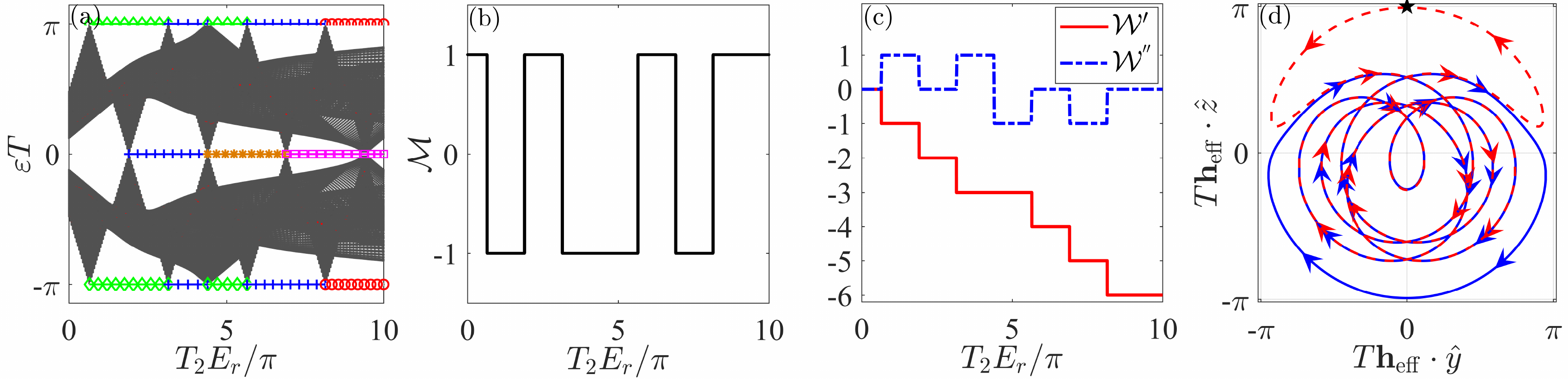}
 \caption{ (a) Quasienergy spectrum, (b) Majorana number, and (c) winding number with the change of $T_2$ for the 1D case. The green $\lozenge$, blue $+$, red $\circ$, brown $*$, and pink $\Box$ symbols denote the cases of one, two, three, four, and six degenerate edge modes, respectively. (d) Trajectories of $\mathbf{h}'_{\rm eff}(k)$ for the duration of $T$ (red dashed lines) and $T+\Delta T$ (blue solid lines) by crossing the phase boundaries of $k=\pi$ by $\bigstar$ for $m_z-2e^{i\alpha}v_0>0$ with $(T_2,T_2+\Delta T)E_r/\pi=(8.14,8.16)$. Other  parameters are $m_z=0.5E_r$, $v_0=0.15E_r$, $v_{so}=0.25E_r$, and $T_1=0.6\pi/E_r$. }
\label{sm1}
\end{figure*}
\section{Recovered chiral symmetry in the 1D case when $T_1\neq T_2$}\label{appe1D}
In this appendix, we show that the chiral symmetry in the 1D case when $T_1\neq T_2$ can be recovered by a unitary transformation. This supplies another subtle way to engineer the large topological number in systems belonging to symmetry class D with $Z_2$ topological invariant.

The evolution operator is $\hat{U}_T=e^{-i\hat{H}_2 T_2}e^{-i\hat{H}_1 T_1}$, where $\hat{H}_j=\mathbf{h}_j(k)\cdot\pmb{\sigma}$ ($j=1,2$). A unitary transformation $\hat{G}_1=e^{i\hat{H}_1T_1/2}$ converts it to $\hat{U}_T'=\hat{U}'_1\hat{U}'_2$ with $\hat{U}'_1=e^{-i\hat{H}_1T_1/2}e^{-i\hat{H}_2T_2/2}$ and $\hat{U}'_2=e^{-i\hat{H}_2 T_2/2}e^{-i\hat{H}_1 T_1/2}$. According to Eq. \eqref{varepsilon}, we have $\hat{U}'_j=\varepsilon'_j I_{2\times 2}+i\mathbf{r}'_j\cdot\pmb{\sigma}$ with $\varepsilon'_1=\varepsilon'_2$ and
$\mathbf{r}'_j=(-1)^j a \underline{\mathbf{h}}_1\times\underline{\mathbf{h}}_2-b\underline{\mathbf{h}}_2-c\underline{\mathbf{h}}_1$,
where $a=\sin |T_1\underline{\mathbf{h}}_1/2|\sin |T_2\underline{\mathbf{h}}_2/2|$, $b=\cos |T_1\underline{\mathbf{h}}_1/2|\sin |T_2\underline{\mathbf{h}}_2/2|$, and $c=\cos |T_2\underline{\mathbf{h}}_2/2|\sin |T_1\underline{\mathbf{h}}_1/2|$. Then we can obtain
$\hat{U}'_T=\varepsilon' I_{2\times 2}+i\mathbf{r}'\cdot\pmb{\sigma}$ with $\varepsilon'=(\varepsilon'_1 )^2-\mathbf{r}'_1\cdot\mathbf{r}'_2$ and
\begin{eqnarray}
\mathbf{r}'&=&2\underline{\mathbf{h}}_1(-\varepsilon'_1 c+ac\underline{\mathbf{h}}_1\cdot\underline{\mathbf{h}}_2+ab)\nonumber\\
&&-2\underline{\mathbf{h}}_2(\varepsilon'_1 b+ac+ab\underline{\mathbf{h}}_1\cdot\underline{\mathbf{h}}_2).\label{r'}
\end{eqnarray}
Equation~\eqref{r'} reveals that if $\mathbf{h}_1(k)$ and $\mathbf{h}_2(k)$ have the same symmetry with the same symmetry operator, then $\hat{U}'_T$ would inherit their symmetry. The similar result can be obtained by $\hat{G}_2=e^{i\hat{H}_2T_2/2}$, which converts $\hat{U}_T$ into $\hat{U}_T''=\hat{U}'_2\hat{U}'_1$.

Consider that both of $\mathbf{h}_j(k)$ in the 1D case possess the chiral symmetry. Choosing $T_1\neq T_2$, the symmetry in $\mathbf{h}_\text{eff}(k)$ determined by $\hat{U}_T$ would be broken. Its topological property is characterized by the Majorana number $\mathcal{M}\equiv\text{sgn}[\mathbf{h}_\text{eff}(0)\cdot\mathbf{h}_\text{eff}(\pi)]_z$ \cite{PhysRevB.90.125143,PhysRevLett.106.220402}. We readily find \begin{equation}\mathcal{M}=\text{sgn}[\sin(T(m_z+2v_0))\sin(T(m_z-2v_0))].\end{equation}
Following the discussion above, the unitary transformation $\hat{G}_j$ makes $\mathbf{h}'_\text{eff}(k)$ preserve the chiral symmetry of $\mathbf{h}_j(k)$. Thus, its topological property is characterized by the winding number. This gives us another way to realize large topological numbers. It has been proven that the number of $0$- and $\pi/T$-mode edge modes relates to the winding number $\mathcal{W}'$ determined by $\hat{U}_T'$ and $\mathcal{W}''$ determined by $\hat{U}_T''$ as \cite{PhysRevB.90.125143}
\begin{equation}
N_0={|\mathcal{W}'+\mathcal{W}''|\over 2},~N_{\pi/T}={|\mathcal{W}'-\mathcal{W}''|\over 2}.\label{nmed}
\end{equation}

We plot in Figs.~\ref{sm1}(a)-\ref{sm1}(c) the quasienergy spectrum, the Majorana number $\mathcal{M}$ determined by $\hat{U}_T$, the winding number $\mathcal{W}'$ determined by $\hat{U}'_T$, and $\mathcal{W}''$ determined by $\hat{U}''_T$ with the change of $T_2$, respectively. With increasing $T_2$, the gap is closed and reopened at $T_2E_r/\pi=0.65$, $3.15$, $4.4$, $5.65$, and $8.15$ for the quasienergy $\pi/T$, and $T_2E_r/\pi=1.9$, $4.4$, $6.9$, and $9.4$ for the quasienergy $0$. They are clearly reflected by $\mathcal{M}$, $\mathcal{W}'$, and $\mathcal{W}''$. However, $\mathcal{M}$ only characterizes the parity of the numbers of the formed edge modes ($\mathcal{M}=-1$ for odd pairs and $\mathcal{M}=1$ for even pairs), while $N_0$ and $N_{\pi/T}$ obtained by recombining of $\mathcal{W}'$ and $\mathcal{W}''$ according to Eq. \eqref{nmed} equal exactly to the numbers of the zero- and $\pi/T$-mode edge modes. Therefore, via the unitary transformations, we have perfectly recovered the bulk-edge correspondence in the $Z_2$ system.  In Fig. \ref{sm1}(d), we show the trajectory of $\mathbf{h}'_\text{eff}(k)$ crossing the quasienergy $\pi/T$ for the band-touching point $k=\pi$. According to Eq. \eqref{zerocon}, $\mathbf{h}'_{\rm eff}(k)$ for $m_z+2v_0>0$ crosses $\pi/T$ along the $-y$ direction. The corresponding $\mathcal{W}$ changes from $-5$ to $-6$. This verifies again our result for the changing rule of the topological number.

\begin{figure*}
 \includegraphics[width=2\columnwidth]{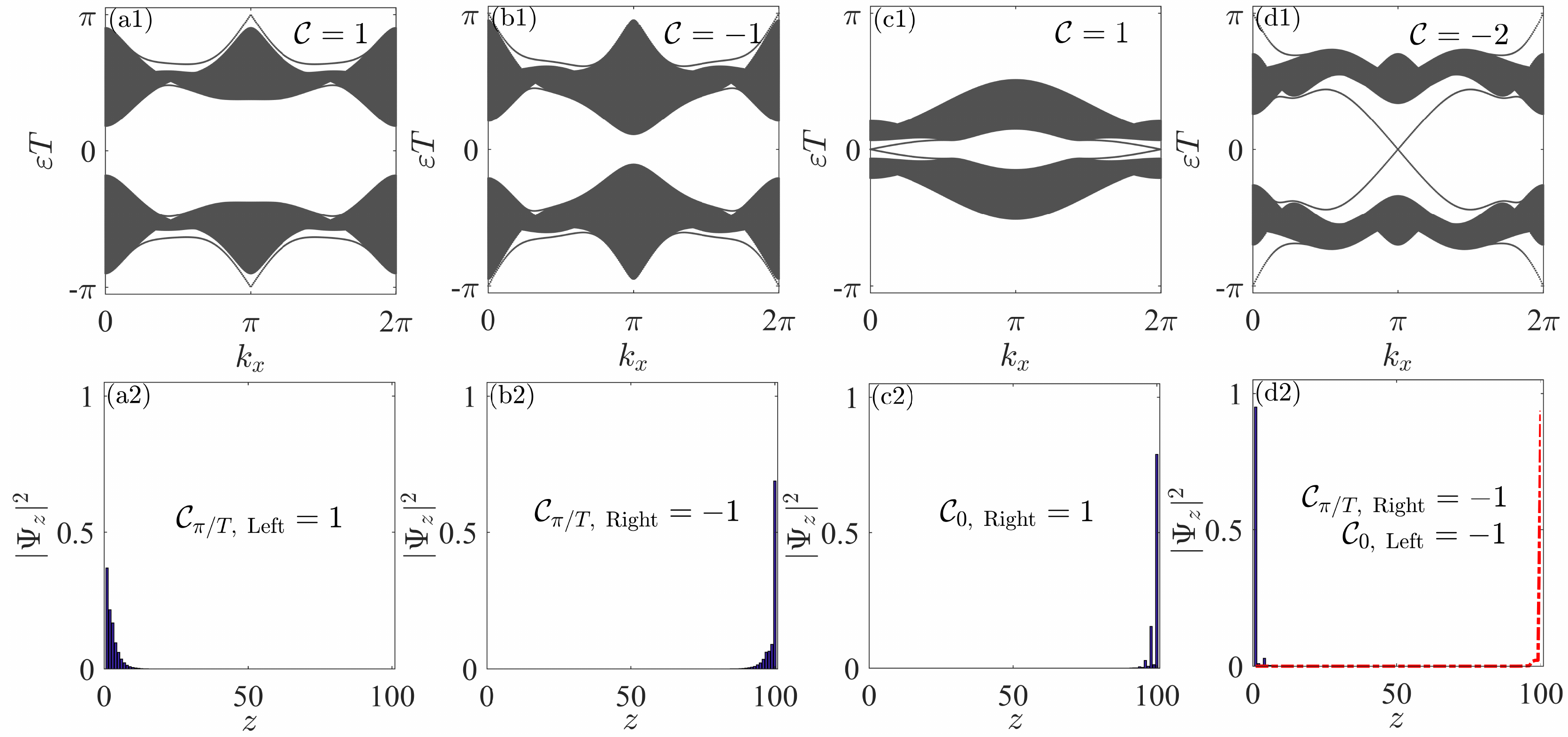}
 \caption{(a1-d1) Quasienergy spectra and (a2-d2) lattice distribution of the edge modes with positive group velocity $\partial_{k_x}\varepsilon$ for the 2D case. The parameters $(TE_r/\pi,v_0/E_r)$ are $(1.8,0.1)$ in panel (a), $(2.1,0.1)$ in panel (b), $(0.3,0.3)$ in panel (c), and $(2.6,0.1)$ in panel (d). The solid and dashed lines in (d2) denote the $0$- and $\pi/T$-mode states, respectively. Other parameters are $m_z=0.5E_r$, $v_{so}=0.28E_r$, and $T_1=0.7T$. }
\label{sm2}
\end{figure*}
\section{Phase transition condition and bulk-edge correspondence in the 2D case}\label{appe2D}

In this appendix, we give the derivation of phase transition condition and the change rule of the topological number in the periodically quenched 2D system.

According to Eqs. (\ref{h1h2})-(\ref{scond}), the boundaries of the quasienergy band closing are determined as follows.

\textbf{Case I}: $T_j|\mathbf{h}_j(\mathbf{k})|=n_j\pi$. In the neighbourhood of the band-touching point satisfying this condition, i.e., $T_j|\mathbf{h}_j(\mathbf{k})|=n_j\pi-\epsilon_j$ with $\epsilon_j$ being an infinitesimal, we obtain $\mathbf{h}_{\rm eff}(\mathbf{k})=(-1)^{n_1+n_2+1}[\epsilon_2\underline{\mathbf{h}}_2(\mathbf{k})+\epsilon_1\underline{\mathbf{h}}_1(\mathbf{k})]/T_{\epsilon_1,\epsilon_2}$. Then we have
\begin{eqnarray}
\lim_{\epsilon_j\rightarrow 0}\partial_{p}\mathbf{h}_{\rm eff}(\mathbf{k})&=&\lim_{\epsilon_j\rightarrow 0}\frac{(-1)^{n_1+n_2+1}}{T_{\epsilon_1,\epsilon_2}}[\epsilon_2\partial_{p}\underline{\mathbf{h}}_2(\mathbf{k})\nonumber\\
&&+\epsilon_1\partial_{p}\underline{\mathbf{h}}_1(\mathbf{k})]=0,
\end{eqnarray}
where $p=k_x$ or $k_y$. It indicates that at the direction of phase transition, the chirality $\text{Ch}(\mathbf{k})$ is zero. Thus, this case has no contribution to the phase transition.

\textbf{Case II}: $\underline{\mathbf{h}}_1(\mathbf{k})=-\underline{\mathbf{h}}_2(\mathbf{k})$. Equation \eqref{h1h2} determines that the bands touch at $\mathbf{k}_0=(k_x,0,k_z)$ satisfying $(\cos k_x +\cos k_z)=\frac{m_z}{2v_0}$. Substituting this condition into Eq. \eqref{bdt}, we obtain $(T_1-T_2)2v_{so}\sqrt{\sin^2k_x+\sin^2k_z}=n\pi$. In the neighborhood of $\mathbf{k}_0$, i.e., $(T_1-T_2)2v_{so}\sqrt{\sin^2k_x+\sin^2k_z}=n\pi-\epsilon$, we have
\begin{widetext}
\begin{eqnarray}
\lim_{\epsilon\rightarrow 0}\partial_{ k_x} \mathbf{h}_{\rm eff} (\mathbf{k})\big|_{\mathbf{k}\rightarrow\mathbf{k}_0}&=&\frac{8 v_0v_{so}\sin |T_1\mathbf{h}_1(\mathbf{k})|\sin|T_2\mathbf{h}_1(\mathbf{k})|\sin k_x}{T|\mathbf{h}_1(\mathbf{k})|^2}[\sin k_x,-\sin k_z,\frac{|\mathbf{h}_1(\mathbf{k})|\sin |T\mathbf{h}_1(\mathbf{k})|}{4v_{so}\sin |T_1\mathbf{h}_1(\mathbf{k})|\sin|T_2\mathbf{h}_1(\mathbf{k})|}],~~~~~~\\
\lim_{\epsilon\rightarrow 0}\partial_{k_z} \mathbf{h}_{\rm eff} (\mathbf{k})\big|_{\mathbf{k}\rightarrow\mathbf{k}_0}&=&\frac{8 v_0v_{so}\sin |T_1\mathbf{h}_1(\mathbf{k})|\sin|T_2\mathbf{h}_1(\mathbf{k})|\sin k_z}{T|\mathbf{h}_1(\mathbf{k})|^2}[\sin k_x,-\sin k_z,\frac{|\mathbf{h}_1(\mathbf{k})|\sin |T\mathbf{h}_1(\mathbf{k})|}{4v_{so}\sin |T_1\mathbf{h}_1(\mathbf{k})|\sin|T_2\mathbf{h}_1(\mathbf{k})|}],~~~~~~
\end{eqnarray}\end{widetext}
which lead to $\partial_{k_x}\mathbf{h}_{\rm eff}(\mathbf{k})\times\partial_{k_z}\mathbf{h}_{\rm eff}(\mathbf{k})=0$. Thus this case cannot induce a topological phase transition either.

\textbf{Case III}: $\underline{\mathbf{h}}_1(\mathbf{k})=\underline{\mathbf{h}}_2(\mathbf{k})$. Equation \eqref{h1h2} requires $\mathbf{k}_0=(\alpha,\beta)$ with $\alpha,\beta=0$ or $\pi$. From Eq. \eqref{bdt}, we have
\begin{equation}
n=T|m_z- (2e^{i\alpha}+2e^{i\beta})v_0|/\pi,\label{kk0}
\end{equation}
which determines the phase boundaries.

To reveal how $\mathcal{C}$ changes when $T$ crosses the phase boundaries, we examine the changing rate of $\mathbf{h}_{\rm eff}(\mathbf{k})$ across the quasienergy $0$ or $\pm \pi/T$ at $\mathbf{k}_0$ in both the $k_x$ and $k_z$ directions, i.e., $\lim_{\epsilon\rightarrow 0}\partial{k_x} \mathbf{h}_{\rm eff}(\mathbf{k}){|}_{\mathbf{k}\rightarrow \mathbf{k}_0}$ and $\lim_{\epsilon\rightarrow 0}\partial_{k_z} \mathbf{h}_{\rm eff}(\mathbf{k})|_{\mathbf{k}\rightarrow \mathbf{k}_0}$ at $T_\epsilon=T-\epsilon$. Using $\mathbf{h}_j(\alpha+e^{i\alpha}\delta,\beta)=[0,(-1)^{j}2v_{so}\delta,q]$, $\mathbf{h}_j(\alpha,\beta+e^{i\beta}\delta)=[(-1)^{j}2v_{so}\delta, 0,q]$ with $\delta$ being an infinitesimal, and Eq. \eqref{varepsilon}, we have
\begin{widetext}
\begin{eqnarray}
\mathbf{h}_{\rm eff}(\alpha+e^{i\alpha}\delta,\beta)&=&\frac{1}{T_\epsilon}[\frac{4v_{so}\delta}{q}\sin |T_1q|\sin|T_2q|,\frac{2v_{so}\delta}{|q|}\sin (T_2-T_1)|q|,\text{sgn}(q)\sin |T_\epsilon q|],\label{Heffx}\\
\mathbf{h}_{\rm eff}(\alpha,\beta+e^{i\beta}\delta)&=&\frac{1}{T_\epsilon}[\frac{2v_{so}\delta}{|q|}\sin (T_2-T_1)|q|,\frac{-4v_{so}\delta}{q}\sin |T_1q|\sin|T_2q|,\text{sgn}(q)\sin |T_\epsilon q|],\label{Heffy}
\end{eqnarray}\end{widetext}
where $q=m_z-2v_0(e^{i\alpha}+e^{i\beta})$. Remembering $n\in \mathbb{Z}$, we conclude that the band touching occurs at the quasienergy $\text{sgn} (q)\pi/T$ for odd $n$ and $ 0_{-\text{sgn} (q)}/T$ for even $n$. The changing rates of $\mathbf{h}_{\rm eff}(\mathbf{k})$ at $\mathbf{k}_0$ are
\begin{widetext}
\begin{eqnarray}
\lim_{\epsilon\rightarrow 0}\partial_{k_x} \mathbf{h}_{\rm eff}(\mathbf{k})|_{\mathbf{k}\rightarrow \mathbf{k}_0}&=&\frac{4v_{so}e^{i\alpha}}{qT}[\sin |T_1 q|\sin |T_2 q|,\frac{\sin(T_2-T_1)q}{2},0],\\
\lim_{\epsilon\rightarrow 0}\partial_{k_z} \mathbf{h}_{\rm eff}(\mathbf{k})|_{\mathbf{k}\rightarrow \mathbf{k}_0}&=&\frac{4v_{so}e^{i\beta}}{qT}[\frac{\sin(T_2-T_1)q}{2},-\sin| T_1 q|\sin |T_2 q|,0].
\end{eqnarray}\end{widetext}
Then $\partial_{k_x}\mathbf{h}_{\rm eff}(\mathbf{k})\times\partial_{k_z}\mathbf{h}_{\rm eff}(\mathbf{k})|_{\mathbf{k}\rightarrow(\alpha,\beta)}=-\frac{16v_{so}^2}{q^2T^2}e^{i(\alpha+\beta)}[0,0,\sin^2 |T_1q|\sin^2|T_2q|+\frac{\sin^2(T_2-T_1)q}{4}]$. The chirality of band-touching points can be calculated as $\text{Ch}[\mathbf{k}=(0,0)]=\text{Ch}[\mathbf{k}=(\pi,\pi)]=-1$ and $\text{Ch}[\mathbf{k}=(0,\pi)]=\text{Ch}[\mathbf{k}=(\pi,0)]=+1$. Using Eqs. (\ref{Heffx}) and (\ref{Heffy}), we have
\begin{eqnarray}
\text{sgn}[\mathbf{h}_{\text{eff}}(\alpha,\beta)|_{T+\epsilon}]_z&=&\text{sgn}(q)\text{sgn}[\sin|q(T+\epsilon)|]\nonumber\\
&&=\text{sgn}(q)\text{sgn}[(-1)^n\epsilon'],\\
\text{sgn}[\mathbf{h}_{\text{eff}}(\alpha,\beta)|_{T-\epsilon}]_z&=&\text{sgn}(q)\text{sgn}[\sin|q(T-\epsilon)|]\nonumber\\
&&=\text{sgn}(q)\text{sgn}[(-1)^{n+1}\epsilon'],~~~
\end{eqnarray}where Eq. (\ref{k0}) is used and $\epsilon'=|q|\epsilon$. According to $\Delta \mathcal{C}(\mathbf{k}_0)=\frac{\text{Ch}(\mathbf{k}_0)}{2}[\text{sgn}[\mathbf{h}_{\text{eff}}]_z|_{T+\epsilon}-\text{sgn}[\mathbf{h}_{\text{eff}}]_z|_{T-\epsilon}]$, we obtain
\begin{eqnarray}
\Delta\mathcal{C}[\mathbf{k}=(0,0)]&=&\text{sgn}[(-1)^{n+1}(m_z-4v_0)],\\
\Delta\mathcal{C}[\mathbf{k}=(\pi,\pi)]&=&\text{sgn}[(-1)^{n+1}(m_z+4v_0)],\\
\Delta\mathcal{C}[\mathbf{k}=(0,\pi)]&=&\Delta \mathcal{C}[\mathbf{k}=(\pi,0)]=\text{sgn}[(-1)^{n}m_z].~~~~~
\end{eqnarray}

The bulk-edge correspondence in the 2D case can be revealed from the quasienergy spectra and site distribution of edge modes with positive group velocity $\partial _{k_x}\varepsilon$ (see Fig.~\ref{sm2}). Figure~\ref{sm2}(a1) has only one pair of $\pi/T$-mode edge states. The corresponding $\mathcal{C}=1$ is uniquely contributed by one of the pair with positive group velocity, which resides on the left edge [Fig. \ref{sm2}(a2)]. Thus the $\pi/T$-mode left-edge state contributes $1$ to $\mathcal{C}$. Similarly, Figs.~\ref{sm2}(b1) and \ref{sm2}(b2) indicate that the $\pi/T$-mode right-edge state contributes $-1$ to $\mathcal{C}$, while Figs.~\ref{sm2}(c1) and \ref{sm2}(c2) show that the $0$-mode right-edge state contributes $1$ to $\mathcal{C}$. Figure~\ref{sm2}(d1) has one pair of $0$-mode and one pair of $\pi$-mode edge states. The $\pi$-mode state resides on the right edge and thus contributes $\mathcal{C}_\pi=-1$. The $0$-mode state resides on the left edge and gives $\mathcal{C}_0=-1$. Then the total Chern number $\mathcal{C}=\mathcal{C}_0+\mathcal{C}_\pi=-2$ can be justified.

\bibliography{references}
\end{document}